\documentstyle[12pt]{article} 
 1                                        
\def\vec#1{\mbox{\protect\boldmath $ #1 $}}

\begin{document}                                                              
\begin{center}                                                                
{ \large\bf THE CONSISTENT  NEWTONIAN LIMIT of EINSTEIN'S GRAVITY WITH  a COSMOLOGICAL CONSTANT \\}
\vskip 2cm  
{ Marek Nowakowski \\}                              
Departamento de Fisica, Universidad de los Andes, A.A. 4976,\\
Santafe de Bogota, D.C., Colombia
\end{center}
\vskip .5cm                             



\begin{abstract} 
We derive the `exact' Newtonian limit of general relativity with a positive cosmological constant
$\Lambda$. We point out that in contrast to the case with 
$\Lambda  = 0 $, the presence of a 
positive $\Lambda$ in Einsteins's equations enforces, 
via the condition  $\vert \Phi \vert
\ll 1$ on the potential $\Phi$, a 
range ${\cal R}_{max}(\Lambda) \gg r \gg {\cal R}_{min} (\Lambda)$, 
within which 
the Newtonian limit is valid. It also leads to the existence of a 
maximum mass, ${\cal M}_{max}(\Lambda)$. 
As a consequence we cannot put the boundary condition for 
the solution of the Poisson
equation at infinity. A boundary condition suitably chosen now at a 
finite range will then get
reflected in the solution of $\Phi$ provided the mass distribution 
is not spherically symmetric. 
\end{abstract}                                                                

\section{Introduction}
The cosmological constant $\Lambda$ can, in principle, enter the 
gravitational equations determining the metric
$g_{\mu \nu}$. It has been introduced by Einstein to save the 
universe from expanding, and rejected by him after
expansion had been discovered by Hubble. Historically, the next stage 
of development regarding this constant can be 
characterized as the search for a mechanism which would allow to put 
the effective cosmological constant,
being the sum of contributions from vacuum fluctuations and a constant
from gravity, to zero \cite{weinberg1} (for other reviews see \cite{carroll} and \cite{starob}).     
Such an explanation is 
desirable to avoid fine-tuning problems.
No generally accepted mechanism has been found. However, something unexpected 
emerged from the recent measurement of
cosmological parameters whose values seem to favour an open 
accelerating universe \cite{newuniv, turner}. 
One of the theoretical
models, in agreement with such a scenario, is Einstein's gravity 
with a positive cosmological constant
\cite{newuniv,turner,narlikar}.
The theoretical efforts shifted then to finding an explanation 
for the actual value of $\Lambda$. One approach invokes the anthropic
principle \cite{barrow} i.e. by considering an ensemble of 
possible universes (be it by inflation 
\cite{inflation}
or quantum cosmology \cite{quantum}) with different cosmological 
constants; the latter characterized by probability
distributions \cite{weinberg2}.
 
Since for a long time the bias has been towards $\Lambda =0$ and, if at all, 
the applications of models containing a non-zero
$ \Lambda$ were thought to be important only at cosmological scales, 
not much attention was paid to the Newtonian limit
in the presence of $\Lambda$. We will show, however, 
that some new results of the Newtonian limit with $\Lambda $ being 
non-zero can be derived, and
that given the presently favoured value of $\Lambda$, 
the vacuum force induced by it is non-negligible at astrophysical 
scale of one Mpc and bigger. 
Hence astrophysical
applications of such a vacuum force are not excluded 
(indeed we will show examples where it can be quite important).
A  careful re-examination of the Newtonian limit, especially when 
it reveals some new insights, is 
therefore welcome.

The Newtonian limit of Einsteins's gravity and its generalizations 
is given in the form of Poisson equation for the 
potential $\Phi$ connected with the metric by the equation
\begin{equation} \label{geodesic}
g_{00} \simeq -(1+2 \Phi),
\end{equation}
which follows from the equation of geodesics for a weak static field 
produced by non-relativistic mass
distribution.
Needless to say, the Poisson equation has to be supplemented by 
some boundary condition which in the simplest case,
general relativity with zero cosmological constant, is more or 
less obvious. This is partly so, because
we know of course what to expect from a Newtonian theory, it 
being chronologically before Einsteins's gravity.
In general, however, we have to make a decision about the kind of 
boundary condition (i.e. Dirichlet, von Neumann
or mixed), where in space we want to put our boundary condition and 
what value the boundary condition is
supposed to assume.  Obviously with $\Lambda =0$ we have Dirichlet 
boundary conditions which we implement as
$\Phi\vert_R = const$ with $R \to \infty$. Note that we can choose 
safely the point $R \to \infty$ as it is
indeed mathematically consistent (i.e. there is no other information 
following from the Newtonian limit which would 
contradict such a choice) as long as we keep in mind that from 
the physics point of view we should not extend
the validity of the Newtonian limit to a scale of the size of the whole 
universe \cite{remark1}. 
Another point worth stressing
is that general relativity offers yet another information on the 
Newtonian potential $\Phi$. It is the 
Schwarzschild solution from which we know, via equation (\ref{geodesic}), the potential already for a
point-like and/or spherically symmetric object. 
This is a valuable information since we then also know, 
without actually solving the Poisson equation,  what 
the potential of any object will look like from far enough distances. We can use this as the value
of the boundary condition at, say, the point $R$ 
with $R \to \infty$ in case $\Lambda =0$ as we said before,
but not so in the case $\Lambda > 0$ as we will show below. 

All these simple and rather obvious considerations might 
change in more sophisticated cases.
It is a curious fact that this already happens  when a positive 
cosmological constant is switched on. What
remains is the need for a suitable boundary condition which, one might be 
inclined to think, could be chosen
also at $R \to \infty$ regardless of the fact that the potential assumes 
an infinite value there (we can treat
this infinity as a constant and subtract it).  But, as we will demonstrate 
in this work, the very same
condition for the Newtonian approximation to hold, namely $\vert \Phi \vert \ll 1$ implies now also
a restriction on the range of the distance $r$. In other words, there exist now a ${\cal R}_{max}$ and
${\cal R}_{min}$ dependent on $\Lambda$ such that
\begin{equation} \label{minmax}
{\cal R}_{min}(\Lambda) \ll r \ll {\cal R}_{max}(\Lambda) .
\end{equation} 
Therefore, it would be mathematically inconsistent to put our boundary 
condition at infinity.
Similarly, the same requirement ($\vert \Phi \vert \ll 1$) leads also to 
the existence of a maximum mass
${\cal M}_{max}$.
Any mass $M$ considered in the Newtonian limit with $\Lambda > 0 $ has to satisfy
\begin{equation} \label{maxmass}
M \ll {\cal M}_{max}(\Lambda)
\end{equation}
Both inequalities, (\ref{minmax}) and (\ref{maxmass}), follow strictly 
from $\vert \Phi \vert \ll 1$.  
  
\setcounter{equation}{0}

\section{The Newtonian limit of Einstein's 
equations with $\Lambda \neq 0$}
The weak field expansion in gravity starts by expressing the 
metric $g_{\mu \nu}$ through the Minkowski metric 
$\eta_{\mu \nu}$ and  spin-2 field $h_{\mu \nu}$ 
(up to a multiplicative constant). 
\begin{equation} \label{hmunu}
g_{\alpha \beta}=\eta_{\alpha \beta} + h_{\alpha \beta}
\end{equation}
For a weak field we have
\begin{equation} \label{NLh} 
\vert h_{\alpha \beta} \vert \ll 1
\end{equation}
This condition translates via (\ref{geodesic}) into
\begin{equation} \label{NL}
\vert \Phi \vert \ll 1
\end{equation}
In this limit Einstein's equation with $\Lambda \neq 0$ (we are using 
the conventions of \cite{weinberg3})
\begin{equation} \label{einstein}
{\cal R}_{\mu \nu} -{1 \over 2} {\cal R} g_{\mu \nu} - 
\Lambda g_{\mu \nu} = -8 \pi G_N T_{\mu \nu}
\end{equation}
reduces, in the first approximation, to the Poisson equation of the form 
\cite{remark0}
\begin{equation} \label{poisson}
\triangle \Phi = (4 \pi G_N) \rho -\Lambda
\end{equation}
where $\rho$ is the mass density function. 

We mentioned already in the introduction that there is yet another
information on $\Phi$. It is contained explicitly in the Schwarzschild 
solution, now with a non-zero 
cosmological constant. Hence for point-like and/or spherically
symmetric objects we deduce using
(\ref{geodesic}) that (see e.g. \cite{adler})
\begin{equation} \label{pointlike}
\Phi (r)= -{G_N M \over r} - { 1 \over 6}\Lambda r^2
\end{equation}
which is valid outside the mass distribution.
To examine the consequences of (\ref{NL}), it suffices for the moment to use $\Phi$ as given in
the last equation. We then have
\begin{equation} \label{NL2}
G_N M \ll d(r) \equiv r -{ 1 \over 6}\Lambda r^3
\end{equation}
Several important conclusions can be drawn from (\ref{NL2}). The very first one follows from the fact
that we are only interested in values of $r$ bigger than or equal to 
zero and that the function $d(r)$ has 
a local maximum at
\begin{equation} \label{localmax}
r_{+}=\sqrt{{2 \over \Lambda}}
\end{equation}
Hence a necessary condition to satisfy (\ref{NL2}) is 
\begin{equation} \label{maxmass2}
{\cal M}_{max}(\Lambda) \equiv {2\sqrt{2} \over 3}{1 \over G_N\sqrt{\Lambda}} \gg M
\end{equation}
which is the explicit version of (\ref{maxmass}) announced in the Introduction. It follows that there 
exists  a restriction
on the absolute value of a mass which can be used in the Newtonian limit.

Further insight, now into the restrictions on the distance $r$ entering 
the Newtonian limit, can be obtained
by considering the equation $G_NM=d(r)$ whose explicit form reads
\begin{equation} \label{cubic}
r^3 -\left({6 \over \Lambda}\right)r + 6{G_N M \over \Lambda}= 0
\end{equation}
On account of (\ref{maxmass2}) there will be three real solutions of 
this equation, one negative and the 
other two positive. Consequently the discriminant $D$ of the cubic 
equation (\ref{cubic}) is negative:
\begin{equation} \label{disriminant}
D= {1 \over \Lambda^2}\left[9(G_NM)^2 -{8 \over \Lambda}\right] < 0
\end{equation}
 which, of course, is nothing else but equation (\ref{maxmass2}) in a 
weaker form, $M < {\cal M}_{max}$.
The fact that $D < 0$ allows us now to find the three solutions using an 
auxiliary angle $\sigma$ or
$\sigma_{{ }_{0}}$ given by
\begin{equation} \label{sigma}
\cos \sigma=\cos\left(\sigma_{{ }_{0}} + {\pi \over 2}\right)=
{ M \over {\cal M}_{max}}
\end{equation}
In view of (\ref{maxmass2}) we can actually write
\begin{equation} \label{phi0}
\sigma_{{ }_{0}} \simeq -{M \over {\cal M }_{max}}
\end{equation}
We then obtain for the three solutions the following parameterized expressions
\begin{eqnarray} \label{solutions}
R_0&=&-2r_{+}\cos\left({\sigma_{{ }_{0}} \over 3} +{\pi \over 6}\right) \nonumber \\
R_1&=&-2r_{+}\cos\left({\sigma_{{ }_{0}} \over 3} +{5\pi \over 6}\right) \nonumber \\
R_2&=&-2r_{+}\cos\left({\sigma_{{ }_{0}} \over 3} +{3\pi \over 2}\right) 
\end{eqnarray}
where $r_{+}$ is given in (\ref{localmax}). One can easily check that $R_0 < 0$ and $R_{1,2} > 0$ as well
as $R_1 >R_2$. Obviously we can now identify $R_1$ with ${\cal R}_{max}(\Lambda)$ and $R_2$ with
${\cal R}_{min}(\Lambda)$ in equation (\ref{minmax}).  More explicitly we obtain
\begin{equation} \label{minmax2}
{\cal R}_{max}\simeq \sqrt{6 \over \Lambda}\left[1 - {1 \over 3\sqrt{3}}{M 
\over{\cal M}_{max}}
\right ] \gg r \gg{\cal R}_{min}\simeq G_NM\left[1 -{1 \over 54}\left({M 
\over{\cal M}_{max}}\right)^2\right]
\end{equation}
where in the expansion for ${\cal R}_{max, min}$ we kept only the first corrections to the leading terms.  
Of course, the right-hand side of this inequality is well known. We keep it here only for completeness
and to display corrections due to non-zero $\Lambda$.

To compare the curious fact given by (\ref{minmax2})
which puts an upper and lower bound on possible
distances used in the Newtonian limit, let us briefly repeat the steps for the case $\Lambda < 0$.
Since in this case we pick up a relative minus between the two terms in (\ref{pointlike}),
we have to introduce a third scale defined by
\begin{equation} \label{thirdscale}
\bar{{\cal R}} \equiv \left({6G_N M \over \vert \Lambda \vert} \right)^{1/3}
\end{equation}
Then in analogy to (\ref{NL2}), we have, 
\begin{eqnarray} \label{NL3}
G_NM \ll \tilde{d}(r) &\equiv& r + {1 \over 6}\vert \Lambda \vert r^3
\,\,\,\, {\rm if} \,\,\,\, \bar{{\cal R}} > r \nonumber \\
G_NM \ll d(r) &=& r - {1 \over 6}\vert \Lambda \vert r^3
\,\,\,\, {\rm if} \,\,\,\, \bar{{\cal R}} < r 
\end{eqnarray}
Let us first concentrate on the first case i.e. we assume that we are with our
$r$ below the value $\bar{{\cal R}}$.
The function $\tilde{d}(r)$ does not have a local maximum as it was
the case with $d(r)$ in (\ref{NL2}).  No restriction on the value of
the mass follows (the use of ${\cal M}_{max}(\vert \Lambda \vert)$ below is only for
algebraic convenience). The solutions of the corresponding cubic
equation give us only one real solution which can be identified with
$\tilde{{\cal R}}_{min}(\Lambda)$. The discriminant is now positive
and correspondingly our parameterized solution is
\begin{eqnarray} \label{solution2a}
\tilde{{\cal R}}_{min}(\Lambda)={2 \sqrt{2} \over \sqrt{\vert 
\Lambda \vert}}\sinh{\tilde{\sigma} \over 3} \nonumber \\
\sinh\tilde{\sigma}\simeq \tilde{\sigma}={ M \over {\cal M}_{max}}
\end{eqnarray}
where in the last equation we {\it assumed}, strictly speaking, that
$M/{\cal M }_{max} \ll 1$ as this does not follow stringently anymore.
This assumption we will keep in the following for simplicity.
It implies in particular that
\begin{eqnarray} \label{byassumption}
\bar{{\cal R}} \gg &\tilde{{\cal R}}_{min}& \simeq {\cal R}_{min}(\vert \Lambda \vert)
\nonumber \\
{\cal R}_{max}(\vert \Lambda \vert) \gg &\bar{{\cal R}}&
\end{eqnarray}
Our restriction on the distance $r$ reads in the case $\Lambda < 0$ as
\begin{equation} \label{min}
r \gg \tilde{{\cal R}}_{min} \simeq G_NM\left[1 + 
{M \over 2 {\cal M}_{max}(\vert \Lambda \vert)}\right]
\end{equation}
which up to small corrections is the same as in the case with zero
cosmological constant, provided that we are with our $r$ below $\bar{{\cal R}}$.  
A priori, however, there could exist universes
with $\Lambda <0$ for which the crucial mass ratio $M/{\cal M}_{max}$
need not be small. Then we would have to work with the full solution
in (\ref{solution2a}) without using the expansion of $\sinh
\tilde{\sigma}$. The coincidence which we had before with the $\Lambda
=0 $ case would vanish. 

On the other hand choosing to go beyond $\bar{{\cal R}}$ implies 
that a consistent Newtonian limit is only possible
for 
\begin{equation} \label{min2}
r \ll {\cal R}_{max}(\vert \Lambda \vert)
\end{equation}
Coming back to the situation where $\Lambda > 0$ some comments on the
actual present values of ${\cal M }_{max}$ and ${\cal R}_{min, max}$
are in order. It is convenient to express the value of $\Lambda$
through a constant density called vacuum density $\rho_{vac}$:
\begin{equation} \label{vacdensity}
\Lambda=8\pi G_N\rho_{vac}
\end{equation}
Such a vacuum density is then best compared to the critical density 
of the universe
\begin{equation} \label{rhocrit}
\rho_{crit}=3H_0^2/8\pi G_N
\end{equation}
where $H_0$ is the Hubble constant given by
$H_0=100h_0 {\rm km} {\rm s}^{-1}{\rm Mpc}^{-1}$. We can then write
\begin{eqnarray} \label{values}
{\cal M}_{max}&=&3.41 \times
10^{22}h_0^{-1}M_{\odot}\left({\rho_{crit}  
\over \rho_{vac}}\right)^{1/2}
\nonumber \\
{\cal R}_{max}&=&4.24 \times 10^{3}{\rm Mpc}
\left[h_0^{-1}\left({\rho_{crit}
  \over \rho_{vac}}\right)^{1/2}
-5.64 \times 10^{-24}\left({M \over M_{\odot}}\right)\right]
\nonumber \\
{\cal R}_{min}&=&1.48 {\rm km} \left({M \over M_{\odot}}\right)
\end{eqnarray}
where we have neglected corrections of the order $(M/{\cal
M}_{max})^2$. With the recent measurements of crucial cosmological
parameters \cite{newuniv}, one of the favoured models in agreement
with these measurements is an open universe with $\Lambda > 0$. The
actual preferred value is $\rho_{vac} \simeq (0.7-0.8) \rho_{crit}$
\cite{turner}. Then equation (\ref{minmax2}) together with
(\ref{values}) can be interpreted in two ways: (i) Even in a universe
void of matter, but with $\Lambda > 0$, the validity of the Newtonian
limit for a test mass $M$ would be restricted by ${\cal M}_{max}$ and
${\cal R}_{max, min}$. As it happens ${\cal R}_{max}$ is close to the
size of the universe at present and ${\cal M}_{max}$ to its mass. (ii)
We could then equally well say, that the restrictions on the Newtonian
limit in form of equations (\ref{minmax2}) and (\ref{values}) taken
with $\rho_{vac} \sim \rho_{crit}$ simply tell us that we should not
apply the Newtonian limit to the whole universe.  This actually does
not follow from $\vert \Phi \vert \ll 1$ when we take $\Lambda =0$ as
noted already in \cite{remark1}. Nevertheless the last two points
remain curious facts concerning, say, any astrophysical applications
at present in the universe we are living in. We emphasize that this is
so, because of the small value of $\rho_{vac}$. This need not be so in
other realistic investigations: (iii) We mention here the possibility
of a time dependent $\Lambda$ which could have been larger in the past
\cite{quintessence}.  (iv) An anthropic principle, modeling other
possible universes (``real'' or hypothetical) would necessarily have
to rely on (\ref{minmax2}) when examining via the Newtonian limit  
structure formation in
universes with a sizeable positive cosmological constant. (v)
Extrapolating the fate of an open universe \cite{fate} with
$\rho_{vac} \sim \rho_{crit}$ (as our own universe seems to be) into a
far future with ${\cal R}_{univ.} \gg {\cal R}_{max}$ the restricting
equations (\ref{minmax2}) and (\ref{values}) have e.g. the virtue to
tell us that even for a diluted conglomeration of clusters of the size
${\cal R}_{max}$ the Newtonian limit is not applicable. It is then
legitimate to ask whether the fact that at present ${\cal R}_{univ.}
\simeq {\cal R}_{max}$ is merely a coincidence.

\setcounter{equation}{0}
\section{Boundary condition and the solution of the Poisson equation}

We choose Dirichlet boundary conditions setting the potential to a
constant value at a distance $R$.  This is almost as in the standard
case with $\Lambda = 0$ save for the fact that in view of
(\ref{minmax2}) we cannot let the distance $R$ to go to
infinity. Hence some finite effects of this finite value of $R$ are to
be expected in the solution for $\Phi$. This is unlike the case with
zero cosmological constant where the boundary condition $\Phi
(\vec{x}) \vert_{R \to \infty} = {\rm const}$ does not leave any
$\vec{x}$ dependent terms in the solution $\Phi (\vec{x})$.

The general solution in the case under consideration in a space region
$V$ reads
\begin{eqnarray} \label{generalsolution}
\Phi (\vec{x}) &=& - \int_{V}d\vec{x'} G(\vec{x},\vec{x'})\left[G_N 
\rho(\vec{x'}) -{\Lambda \over 4\pi }\right] \nonumber \\
&-& {1 \over 4 \pi}\int_{\partial V}dS' \Phi \vert_{\partial V} 
\hat{\vec{n}} \cdot 
\vec{\nabla}_{\vec{x'}}G(\vec{x}, \vec{x'})\vert_{\partial V}
\end{eqnarray}
where $\Phi \vert_{\partial V}$ is the Dirichlet boundary condition
chosen at the surface $\partial V$ of the volume $V$. To be more
specific we opt in this case for
\begin{equation} \label{boundary}
\Phi \vert_{\partial V}= \Phi \vert_{R}= 
-{G_N M \over R} - {1 \over 6}\Lambda R^2
\end{equation}
i.e. the potential of a point mass as seen from the large distance
$R$.  $G(\vec{x}, \vec{x'})$ in (\ref{generalsolution}) is the Green's
function which we give here in terms of spherical harmonics.
\begin{eqnarray} \label{green}
G(\vec{x}, \vec{x'}) &=& 4 \pi \sum_{l=0}^{\infty}\sum_{m=-l}^{l}
{Y^{\ast}_{lm}(\theta', \varphi')
Y_{lm}(\theta, \varphi) \over 2l+1} \nonumber \\
& \times & r^l_< \left[{1 \over r_>^{l+1}} -{r_>^l \over R^{2l+1}}\right]
\end{eqnarray}
with
\begin{eqnarray} \label{r}
r_< &=& {\rm min}(\vert \vec{x} \vert , \vert \vec{x'} \vert)
\nonumber \\
r_> &=& {\rm max}(\vert \vec{x} \vert , \vert \vec{x'} \vert)
\end{eqnarray}
Solving the surface integral in (\ref{generalsolution}) and using the identity
\begin{equation} \label{identity}
{1 \over \vert \vec{x} - \vec{x'} \vert}=4\pi 
\sum_{l=0}^{\infty}\sum_{m=-l}^{l}
{1 \over 2l+1}{r_<^l \over r_>^{l+1}}Y^{\ast}_{lm}(\theta', \varphi')
Y_{lm}(\theta, \varphi)
\end{equation}
the solution can be conveniently rewritten as
\begin{eqnarray} \label{solution2}
\Phi(\vec{x}) =&-&G_N \int_{V'}d\vec{x'}{\rho(\vec{x'}) \over \vert
\vec{x} -\vec{x'}\vert}
- {1 \over 6} \Lambda \vert \vec{x}\vert^2 \nonumber \\
&+& G_N\int_{V'}d\vec{x'}G'(\vec{x}, \vec{x'})\rho(\vec{x'}) \nonumber \\
&-& \left({G_NM \over R} + {1 \over 6}\Lambda R^2 \right)
\end{eqnarray}
where  $V' \subset V$ is the space region in which $\rho(\vec{x'}) \neq 0$ and 
\begin{equation} \label{G'}
G'(\vec{x}, \vec{x'}) \equiv 4\pi  
\sum_{l=0}^{\infty}\sum_{m=-l}^{l}{Y^{\ast}_{lm}(\theta', \varphi')
Y_{lm}(\theta, \varphi) \over 2l+1} 
{r^l_< r_>^l \over R^{2l+1}}
\end{equation}
The last two equations are what we would like to call the `exact'
Newtonian limit of Einstein's gravity with a positive cosmological
constant. The first two terms in (\ref{solution2}) are what we would
naively expect were we allowed to put the boundary condition at $R \to
\infty$. The third term represents exactly the effect of the boundary
condition taken at finite distance $R$.  The name `exact' Newtonian
limit simply refers to the fact that in view of (\ref{minmax2})
mathematical consistency forces us to put the boundary condition at a
finite range which in turn implies (\ref{solution2}).  The last term
in (\ref{solution2}) is a constant and as such of no further
importance.

It might appear that we have neglected at the boundary 
(i.e. postulating our boundary condition) terms of the order $1/R^n$ 
and at the same time retain terms of the same order in the third 
term in (\ref{solution2}). This is not a contradiction, however.
Despite the physical appeal of the argument that far away of 
the matter distribution the potential has point-like structure,
our choice of the boundary condition (\ref{boundary}) has been quite arbitrary.
Indeed, we could have opted for any other constant value for $\Phi\vert_R$
without changing the solution (\ref{solution2}), up to an unessential constant.
This shows that actually there is no such contradiction.

Evidently we are now confronted with the question which value of $R$
we should be using in the general solution (\ref{solution2}). This is
not as unique as one might wish, but for any practical purposes we
will offer a solution to this problem. First, with a given mass
distribution $\rho$ and our solution in form of equation
(\ref{solution2}), we could copy some of the steps in the last section
which led us to the existence of ${\cal R}_{max,min}$ and ${\cal
M}_{max}$ and the restricting equations (\ref{minmax2}) and
(\ref{maxmass2}). This is, however, in general not necessary. Note
that the third `new' term in (\ref{solution2}) becomes a constant for
a spherically symmetric body.  This is as it should be since in the
Newtonian limit gravity with a cosmological constant shares with
Newtonian gravity ($\Lambda =0$) the virtue that in both cases a
spherically symmetric object has the same potential as the point-mass
(see e.g. \cite{barrow}). Our solution, of course, reduces up to a
constant to (\ref{pointlike}) for spherical objects.  For bodies which
are not spherically symmetric, new parameters of the dimension of
length describing the off-sphericity of the object will appear. These
parameters will be much smaller than ${\cal R}_{max}$ in
(\ref{minmax2}), independent of the numerical value of the
latter. Then the equations (\ref{maxmass2}) and (\ref{minmax2})
restricting the mass and the distance used in the Newtonian limit hold
also for the general solution (\ref{solution2}). If so, the only
possible large distance is some $R$ which is one or two orders
of magnitude smaller than ${\cal R}_{max}$. Note that 
(\ref{NL}) does not 
allow
us to identify $R={\cal R}_{max}$. 
It is also worth pointing out that, since we correlate $R$ with ${\cal R}_{max}$
(i.e. $R$ is a fraction of ${\cal R}_{max}$), to recover the standard Newtonian result from
(\ref{solution2}) it suffices to take $\Lambda \to 0$ as ${\cal R}_{max} \propto 1/\sqrt{\Lambda}$
which then implies $R \to \infty$.
This seems a reasonable solution so
far. Whether it is stringent remains to be investigated in more detail as
right now a certain amount of ambiguity stays with us. 
We think,
however, for reasons outlined above that (\ref{minmax2}) and
(\ref{maxmass2}) are valid for any mass distribution and its
gravitational potential.

In any case, it is clear that the third term in (\ref{solution2}) will
be suppressed by powers of $1/R$ and therefore normally small. Whether
to retain it in an actual calculation depends, of course, on the
requirements of accuracy of our calculation. The most promising case
we can imagine is a galaxy at the outskirts of some supercluster where
we could hope to see some effects of the boundary condition. Be it as
it may, we think that it is of some importance to point out the
mathematically consistent Newtonian limit of Einstein gravity with
positive cosmological constant.

The discussion of a rather subtle point we have left till the end of this section.
It is about the choice of the origin of the sphere with the radius $R$ at whose boundary
we have put our boundary conditions. First note that the origin of our coordinate system
coincides with the origin of this sphere. This has been done on purpose to simplify the formulae
and is not unlike the coordinate system used in (\ref{pointlike}). 
The choice to put the origin of the coordinate system into the origin of the $R$-sphere is 
also the reason why terms of the form $\vert \vec{x} -\vec{R_0}\vert^2$, 
where $\vec{R_0}$ would be a vector between the two origins, do not appear in 
(\ref{solution2}). Equipped with these remarks we can now make the following statements
a.) the choice of the origin of the $R$-sphere in space (now together with the coordinate system)
is arbitrary as no distinguished point exists and b.) the repulsion of the test 
particle due to the
$\Lambda$ term will be always away from the center of this sphere. Taking these two
statements together we conclude that any imaginary point in space will experience a repulsion
from any other point due to the $\Lambda$ term. We can interpret this as 
a kinematical effect of the 
expansion of the universe (due to $\Lambda$) which will now remain even in the Newtonian limit.
This has been noticed actually long ago in a different context in \cite{bacry} 
which we will briefly touch now. The arbitrariness of the origin of the spheres requires
that there exist a transformation law between two possible choices. 
This transformation is not the 
Galilean one. The reason is that the Minkowski metric 
is not a solution of the Einstein's equations in the presence of $\Lambda$. The symmetry group 
of the tangent space-time is rather the de Sitter group (anti de Sitter if $\Lambda$ negative).
By a non-relativistic group contraction ($c \to \infty$) of the Poincare group we obtain
the Galilean transformation. Similarly now making the same group contraction 
in the de Sitter group we arrive at what is called the Newton group \cite{bacry}. 
The corresponding space-time has been coined Newton-Hooke space-time to distinguish it from the
normal euclidean space with Galilean transformations \cite{pereira}. The transformations
of the Newton-Hooke space-time read now \cite{bacry}
\begin{eqnarray} \label{transformations}
\vec{x'}&=&{\cal U}\vec{x} +\vec{v}\tau\sinh {t \over \tau} +\vec{a}\cosh 
{t \over \tau}\nonumber \\
t'&=&t+b
\end{eqnarray}
where $b$ is time translation, $\vec{a}$ space translation, $\vec{v}$ a constant velocity
and ${\cal U}$ a rotation matrix. The parameter $\tau$ is proportional to $1/\sqrt{\Lambda}$.
One can see from these equations that starting with a displacement $\vec{a}$, one ends up
with a growing distance between the two vectors $\vec{x}$ and $\vec{x'}$ which 
from the point of view of non-relativistic physics is a kinematical effect due to $\Lambda$.
We can therefore state that the interpretation of the two Newtonian limits, one of the
Einsteins's equations and the second one of the symmetry group via group contraction, agree
as it should be since the latter is the symmetry of the former (dynamics).  
\setcounter{equation}{0}
\section{Astrophysical applications}

With the presently favoured value of $\rho_{vac} \sim \rho_{crit}$ one
might be inclined to think that the presence of a positive
cosmological constant is only of relevance for cosmology where we are
not allowed to use the Newtonian limit, but the full set of Einstein's
equations.
However, by examining some simple examples below we can convince
ourselves of just the opposite; the cosmological constant can be of
relevance in astrophysical applications where the Newtonian limit
plays a roles.  We will discuss three examples, one of field galaxies
and the other two for larger structures like galaxy clusters and
superclusters.

(a) Two field galaxies at a far away distance from any cluster could,
in principle, display effects of `anti-gravity' due to the
cosmological constant. Consider the ratio
\begin{equation} \label{fieldgalaxy}
\left({\vert \vec{F}_{vac}\vert \over \vert \vec{F}_{Newton}\vert} 
\right)_{{\rm field \, galaxies}}
= {(\Lambda r/3) \over (G_NM/r^2)}
=2.32 \times 10^{12}h_0^2\left({r \over {\rm
Mpc}}\right)^3\left({M_{\odot}  \over M}\right)\left(
{\rho_{vac} \over \rho_{crit}}\right)
\end{equation}
where $\vec{F}_{vac}$ is the force induced by the cosmological
constant and $\vec{F}_{Newton}$ the standard $1/r^2$ Newtonian
force. To pick up a specific example let us fix $M_{\odot}/M \sim
10^{-11}$ and $ r \sim 1$ Mpc with $\rho_{vac} \sim
0.8\rho_{crit}$. The ratio in (\ref{fieldgalaxy}) becomes
\begin{equation} \label{fieldgalaxy2}
\left({\vert \vec{F}_{vac}\vert \over \vert \vec{F}_{Newton}\vert} 
\right)_{{\rm field \, galaxies}}
\simeq 18.6h_0^2\sim 4.6
\end{equation}
the last estimate with $h_0 \sim 0.5$. Field galaxies less massive and
at a larger distance (than in our example above) would certainly ``repel''
each other (the repulsion is, of course, not due to the galaxies since any two space points will
experience repulsion). 
A thorough survey of such field galaxies and their
possible peculiar velocities would be a worthwhile task.

The example we have been considering is not unlike the system of our
own galaxy and Andromeda.  However, with the difference that the
latter are embedded in the Local Group. For clusters or even bigger
objects a comparison of densities is then more suitable.  (b) Hence,
considering now such clusters \cite{clusterp, bahcall}, and assuming
them to be roughly spherical we can write for the ratio
\begin{equation} \label{cluster}
\left({\vert \vec{F}_{vac}\vert  \over \vert \vec{F}_{Newton}\vert}
  \right)_{{\rm cluster}}= 
{(\Lambda r^3/3) \over (G_N\rho_{cl} V(r))}
=2\left({\rho_{crit} \over \rho_{cl}}\right)\left(
{\rho_{vac} \over \rho_{crit}}\right)
\end{equation}
where $V$ is the volume of the cluster and $\rho_{cl}$ its mass
density.  For a galaxy in the cluster of a typical density $\rho_{cl}
\sim 3 \times 10^{-28}{\rm g\,cm}^{-3}$ the ratio becomes
\begin{equation} \label{cluster2}
\left({\vert \vec{F}_{vac} \vert \over \vert \vec{F}_{Newton}\vert}
 \right)_{{\rm cluster}} \sim 
0.13 \, h_0^2\left(
{\rho_{vac} \over \rho_{crit}}\right) \, \sim 0.025
\end{equation}
with the assumption of $\rho _{vac} = 0.8 \rho _{crit}$. This is
actually not too small especially if `diluted' clusters with density
less than $3 \times 10^{-28} {\rm g\,cm}^{-3}$ exist. It might be
again worthwhile to hunt for such diluted objects. In \cite{bahcall}
very low density clusters of $M=10^{12.5}M_{\odot}$ and radius
$R=1.5$ Mpc are mentioned. This would correspond to a density of
$1.5 \times 10^{-29} {\rm gcm}^{-3}$ and our ratio becomes now
\begin{equation} \label{cluster3}
\left({\vert \vec{F}_{vac} \vert \over \vert \vec{F}_{Newton}\vert}
 \right)_{{\rm cluster}} \sim 0.5
\end{equation}
Certainly, for rich clusters the cosmological constant does not play any role.

(c) It is well known that the densities of superclusters are of the
order of $\rho_{crit}$ or even less \cite{zomby,zeldo}. This, in our
case, is a very important observation.  In this case the `vacuum
force' $\vec{F}_{vac}$ could become important or even dominant for a
galaxy (or cluster) at the edge of such a conglomeration. To be
specific we quote the case for the Local Supercluster below. Using the
values of the mass M $= 5 \times 10^{48} {\rm g}$ and radius R $= 3.16
\times 10^{25}{\rm cm}$ as given in ref. \cite{zomby} for the Local
Supercluster, we get,
\begin{equation} \label{supercluster1}
\left({\vert \vec{F}_{vac} \vert \over \vert 
\vec{F}_{Newton}\vert} \right)_{{\rm Local}} =  
{(\Lambda r^3/3) \over (G_N\rho_{supercl} V(r))}
 = {8 \pi \over 3} \, r^3 \, {\rho _{vac} \over M_{supercl}(r)} = 0.2
\end{equation}
The density of the above supercluster is 4 $\times 10^{-29}{\rm
g\,cm}^{-3}$.  For the numerical estimate we consider a test body
again at the edge of such a supercluster and use the same value for
$\rho_{vac}$ as in (\ref{cluster2}).  We can see that the importance
of the vacuum force already increased a lot for the supercluster over
typical clusters (see eq. \ref{cluster2}) with a density which is an
order of magnitude smaller than that of the cluster in
eq. (\ref{cluster2}).  For superclusters with even smaller densities
like those quoted by the Lick Observatory survey \cite{super2},
namely, $\rho _{supcl} = 2.5 \times 10^{-30}{\rm g\,cm}^{-3}$ and
$\rho _{supcl} = 3.16 \times 10^{-31}{\rm g\,cm}^{-3 }$ in
ref. \cite{super3}, the effect of the vacuum force would be much
larger, of the order
\begin{equation} \label{supercluster2}
\left({\vert \vec{F}_{vac} \vert \over \vert 
\vec{F}_{Newton}\vert} \right)_{{\rm supercluster}} \sim 
2 - 20
\end{equation} 

We have seen that in the Newtonian limit there are realistic chances
that the force induced by a positive cosmological constant plays a
non-negligible role in astrophysics. A detailed survey of pairs of
field galaxies, non-rich (diluted) clusters and superclusters would
indeed help us to shed some light on the cosmological constant.
Questions concerning the applicability of the virial theorem to such
objects are indeed valid questions.  It is not at all clear, e.g.,
whether superclusters represent gravitationally bound systems. We hope
to come to such questions in a more systematic way in a future
publication.

\setcounter{equation}{0}
\section{Conclusions}

The Newtonian limit of Einstein's equations requires that the
gravitational potential should satisfy the strong inequality $\vert
\Phi \vert\ll 1$. We have shown that in the case of a positive
cosmological constant $\Lambda$ this inequality leads to the existence
of an upper bound on the mass (${\cal M}_{max}$) and the distance
(${\cal R}_{max}$) to be used in the Newtonian limit.  Consequently,
we cannot put the boundary condition for the potential at infinity
(which we do if $\Lambda=0$ and can do if $\Lambda < 0$). In the
solution of the Poisson equation there will appear then a term
reflecting this boundary condition at a finite distance. We found
these facts curious enough to merit a note.

With the presently favoured value of $\Lambda$, ${\cal R}_{max}$ and
${\cal M}_{max}$ come very close to the values possessed at present by
our universe. Both maximum values are independent of epoch.  We find
it then a strange coincidence that ${\cal R}_{max} \sim {\cal
R}_{univ}$ and ${\cal M}_{max} \sim {\cal M}_{univ}$. After all, we
could have been living in a universe whose mass and radius are larger
or smaller than the maximum values coming from the analysis of the
Newtonian limit with positive $\Lambda$. The last statement could
equally well be posed as a question.  A closer inspection seems
worthwhile.

We have also pointed out some possible astrophysical application of 
the vacuum force induced by 
$\Lambda$. Possible candidates being inflicted by the $\Lambda$-force, 
range from field galaxies at
a distance of more than one Mpc to clusters and superclusters. Especially for the latter it 
is certainly interesting to revise the virial theorem.

\vskip 1cm

{\bf Acknowledgments}. 
Discussions with N. G. Kelkar, M. Drees, J. V. Narlikar, E. Gorbar, 
Y. Shtanov, R. Rosenfeld, G. Matsas and J. C. Sanabria are gratefully acknowledged.
This work has been partly 
done at Universidad Estadual Paulista supported by 
Funda\c{c}\~ao de Amparo \`a Pesquisa do Estado de S\~ao Paulo (FAPESP)
and Programa de Apoio a N\'ucleos de Excel\^encia (PRONEX).


\begin{thebibliography}{99}
\bibitem{remark1}
In case $\Lambda =0$ we get from $\vert \Phi \vert \ll 1$ the
restriction $r \gg G_N M$. This is not sufficient to say something about
the applicability of the Newtonian limit to the whole universe.
\bibitem{remark0}
Strictly speaking, we have in the weak field approximation in the
harmonic gauge the wave equation
$$\Box h_{\mu \nu}= -16\pi G_N \tilde{S}_{\mu \nu}$$
with
$$ \tilde{S}_{\mu \nu}= \tilde{T}_{\mu \nu}- { 1\over 2}\eta_{\mu \nu}
\tilde{T}^{\lambda}_{\,\,\,\, \lambda}$$ and
$$\tilde{T}_{\mu \nu}=T_{\mu \nu} - {\Lambda \over 8\pi G_N}g_{\mu \nu}$$ 
Assuming now the static case we get for $h_{00}$
$$\triangle h_{00}= -16\pi G_N\tilde{S}_{00}$$
which reduces to (\ref{poisson}) after inserting in the above equation 
$\tilde{S}_{00} \simeq (1/2) \rho -\Lambda/(8\pi G_N)$ and $h_{00}= -2\Phi$
\bibitem{remark3}
The actual usage of the phrase ``repulsion'' due to the cosmological constant might be slightly
misleading as the $\Lambda$-force is not an internal force. What it means is simply that
the galaxies will drift apart. See the discussion at the end of section 3 in this context.
\end{thebibliography}
\end{document}